\documentclass{article}
\usepackage{spconf,amsmath,graphicx}
\usepackage{amssymb}
\usepackage{xcolor}
\usepackage{multirow}
\title{Joint Normality Test via Two-dimensional Projection}
\name{Sara ElBouch and Olivier J. J Michel and Pierre Comon\thanks{This work has been  supported by the MIAI
chair “Environmental issues underground”of Institut MIAI@Grenoble Alpes
(ANR-19-P3IA-0003).}}
\address{Univ. Grenoble Alpes, CNRS, Grenoble INP, GIPSA-Lab, Grenoble, France}
\newcommand{\matr}[1]{\boldsymbol{#1}}

\newcommand{\vect}[1]{\boldsymbol{#1}}
\newcommand{\eqdef}{\stackrel{def}{=}}
\newcommand{\E}[1]{\mathop{\mbox{$\mathbb{E}$}}\{#1\}} % esperance math
\newcommand{\RR}{\mathbb{R}}
\newtheorem{definition}{Definition}[section]
\newtheorem{theorem}{Theorem}[section]

\begin{document}
%\ninept
%
\maketitle
\begin{abstract}
Extensive literature exists on how to test for normality, especially for identically and independently distributed (i.i.d) processes. The case of dependent samples has also been addressed, but only for scalar random processes. For this reason, we have proposed a joint normality test for multivariate time-series, extending Mardia's Kurtosis test. In the continuity of this work, we provide here an original performance study of the latter test applied to two-dimensional projections. By leveraging copula, we conduct a comparative study between the bivariate tests and their scalar counterparts. This simulation study reveals that one-dimensional random projections lead to notably less powerful tests than two-dimensional ones.
\end{abstract}
\begin{keywords}
Multivariate normality test, kurtosis, colored processes, time-series, dependent samples, copula
\end{keywords}
\section{Introduction}
\label{sec:intro}
The popularity of normality, being an underlying assumption to many forecasting and inference models, has led to the development of many procedures aiming at testing the hypothesis of Gaussianity; especially in the case of independent (univariate or multivariate) samples, see the surveys of \cite{Mard80} and  \cite{Henz02:statp}.

Despite the practical importance of having statistically dependent variables, the majority of the tests are derived under the assumption that the latter are identically distributed and \textit{independent}, see \cite{ShapWC68:asaj}, \cite{BowmS75:biom} and \cite{Mard70:biom} to cite few of an extensive literature.  There has been considerable efforts to test the goodness-of-fit of stationary colored processes, such as the Epps test \cite{Epps87:as} based on the characteristic function, the Lobato Velasco's (LV) \cite{LobaV04:et} modification of the test statistic proposed by \cite{BowmS75:biom}, and a test statistic ${}_{k}RP$ that uses 1-D random projections  \cite{NietCG14:csda} to upgrade Epps and LV procedures. 

The lack of testing procedures for dependent samples is exacerbated in the multivariate setting. Available tests are scarce, and a powerful test like the bi-spectrum proposed in \cite{Hini82:jtsa} suffers from severe drawbacks in practice. For this reason, we have recently proposed a computationally efficient test for multivariate time-series \cite{Elbo21}, specifically in the bivariate case. Our work is at the crossroad between all these works: that is, those on multivariate procedures, i.e testing the joint normality and those derived for colored processes.
The questions addressed in this communication are:
\begin{itemize}
    \item In the same spirit as \cite{MalkFA73:jasa, NietCG14:csda} using one-dimensional projection, what can we say about the power of our normality test \cite{Elbo21} when applied to two-dimensional projections?
    \item What is the impact of taking into account the statistical dependence among time samples?
\end{itemize}
 Our main contributions may be summarized as follows: 
 \begin{itemize}
     \item The use of a joint normality test applied to two-dimensional projections of  $p$-variate colored processes.
     \item Copula-based computer experiments confirm that testing two-dimensional random projections is far better than their scalar counterparts applied on one-dimensional projections. This observation is more noticeable for colored processes.
 \end{itemize}
 \textbf{Organisation of the paper.} 
We first formulate the normality test as a binary hypothesis test in  Section \ref{sec:problemform}.
The test statistic \cite{Elbo21} is defined in Sections \ref{sec:scalarprocess} and \ref{sec:bivariate-expressions} ; its asymptotic mean and variance for different scenarios (multivariate) i.i.d and scalar or bivariate colored processes are stated. Sections \ref{sec:computer-experiments} and \ref{sec:mainresults} are dedicated to computer experiments.

\section{Problem formulation}
\label{sec:problemform}
Let $\vect{x}(n) =[x_{1}(n),x_{2}(n),\dots,x_{p}(n)]^{T} \in \RR^{p}$ be a $p$-variate stochastic process. In this paper, the processes are considered zero-mean stationary with finite moments up to order $16$. Let $\matr{S}(\tau)=\E{\vect{x}(n) \vect{x}(n-\tau)^{T}}$ be the co-variance function whose entries are $S_{ab}(\tau)$. Also denote $\matr{S}(0)=\matr{S}$. It is also assumed that $\vect{x}(n)$ is strong-mixing so that the series  $\sum_{\tau=0}^{\infty}S_{ab}(\tau)$ converges absolutely.
The problem is formulated as:
\begin{quote}
\textbf{Problem P1:} Given a sample of size $N$ of $\vect{x}(n)$,
$\matr{X}\eqdef\{\vect{x}(1),\dots,\vect{x}(N)\}$, test
\begin{equation}\label{eq-testGeneral}
    H_0: \matr{X} ~ \text{is Gaussian} \quad versus \quad \bar{H}_0
\end{equation}
where variables $\vect{x}(n)\in \RR^{p}$ are identically distributed, but not statistically independent.
\end{quote}
This normality test belongs to the class of \textit{tests without alternative}. In this framework, a single parameter defines the nominal level of the test:
\begin{equation}\label{eq:levelTest}
    \alpha = \mathbb{P}(\text{choose }\bar{H}_0| H_0 \text{ is true})
\end{equation}
For $p=1$, a standard measure of the gap from normality is the estimated Kurtosis:
\begin{equation}
    \hat{b}_{2} = \frac{1}{N}\frac{\sum_{n=1}^{N} x(n)^{4}}{S^{2}}
\end{equation}
Following Mardia's definition, we assume the extension of this measure to multivariate processes. 
\subsection{Mardia's Kurtosis}
The multivariate counterpart of the empirical kurtosis takes the form of:
\begin{equation}
    B_{p}(N) = \frac{1}{N}\sum_{n=1}^{N}(\vect{x}(n)^{T}\matr{S}^{-1}\vect{x}(n))^2
\end{equation}
with $\matr{S}$ the covariance matrix. Usually, this quantity is unknown and should be estimated on observations.\\
Our final test statistic takes the form:
\begin{equation}\label{estimKurt-eq}
     \hat{B}_p(N) = \frac{1}{N} \sum_{n=1}^{N}\big(\vect{x}(n)^T\widehat{\matr{S}}^{-1}\vect{x}(n)\big)^2
\end{equation}
with
$$
\widehat{\matr{S}}=\frac{1}{N}\sum_{k=1}^{N}\vect{x}(k)\vect{x}(k)^{T}.
$$
\begin{theorem}\cite{Mard70:biom}\label{mk}
Let $\{\vect{x}(n)\}_{1 \leq n \leq N}$ be \textit{i.i.d.} of dimension $p$. Then under the null hypothesis $H_0$, $\hat{B}_p(N)$ is asymptotically normal, with mean $p(p+2)\frac{N-1}{N+1}$ and variance $\frac{8p(p+2)}{N}+o(\frac{1}{N})$.
\end{theorem} 
Thus, we can test  normality by measuring the normalized gap $z$ under $H_0$:  \begin{equation}
    z=(\hat{B}_p(N) - p(p+2))/\sqrt{(8p(p+2)/N)} \sim \mathcal{N}(0,1)
\end{equation}
with $\sim$ means distributed as, and $\mathcal{N}(0,1)$ denotes the univariate standard normal distribution.
We reject the null hypothesis $H_0$ at a significance level $\alpha$ if:
\begin{equation*}
2(1-\Phi(z)) < \alpha
\end{equation*}

where $\Phi$ denotes the cumulative distribution function (cdf) of the standard normal distribution. 
A similar theorem without assuming independence among samples has been devised in \cite{Elbo21} for bivariate statistically dependent processes. For the rest of the paper, Mardia's test statistic will be denoted $\hat{B}_{p,i.i.d}$ to distinguish it from the tests assuming statistical dependence.
\section{Mean and variance of $\hat B_1$ for a scalar colored process}\label{sec:scalarprocess}
In the case of scalar colored samples, the expressions of mean and variance of kurtosis $\hat{B}_1(N)$ are \cite{Elbo21}:
\begin{equation}\label{meanb1}
    \E{\hat{B}_{1}} = 3 - \frac{6}{N} - \frac{12}{N^2} \sum_{\tau=1}^{N-1} (N-\tau) \,\frac{S(\tau)^2}{S^2}  + o(\frac{1}{N})
\end{equation}
\begin{equation}\label{varb1}
    \mathbb{V}\text{ar}\{\hat{B}_{1}\} = \frac{24}{N}\Big[1+\frac{2}{N}\sum_{\tau=1}^{N-1}(N-\tau)  \frac{S(\tau)^{4}}{S^{4}}\Big]   + o(\frac{1}{N}) 
\end{equation}
The dependence between time samples is taken into account in the terms $S(\tau)$. Interestingly, if $S(\tau)=0, \, \forall \tau \neq 0$, the equations above reduce to the i.i.d case.

\noindent
In practice, the time-structure is unknown and the terms $S(\tau)$ are replaced by their sample counterparts.

\section{Mean and variance of $\hat B_2$ in the bivariate case} \label{sec:bivariate-expressions}
In the bivariate case, expressions become rapidly much more complicated but we can still write them explicitly \cite{Elbo21}, as reported below:
\begin{equation}
    \E{\hat{B}_{2}} = 8 - \frac{16}{N} -  \frac{4}{N^2}\sum_{\tau=1}^{N-1} \frac{(N-\tau) Q_{1}(\tau)}{(S_{11}S_{22}-S_{12}^2)^2} + o(\frac{1}{N}) 
\end{equation}
\begin{equation}
    \mathbb{V}\text{ar}\{\hat{B}_{2}\}=\frac{64}{N}+\frac{16}{N^{2}}\sum_{\tau=1}^{N-1}\frac{(N-\tau)Q_{2}(\tau)}{(S_{11}S_{22}-S_{12}^{2})^4} + o(\frac{1}{N}) 
\end{equation}
In the above equations, two kinds of dependence appear: so-called spatial cross-variate dependence $S_{ab}$, and the dependence between time-samples, $Q_i(\tau)$.
Due to their length, expressions of $Q_{1}(\tau)$ and $Q_{2}(\tau)$ are not explicited here and can be found in \cite{Elbo21}.

\noindent
In practice, the time-structure is unknown and the terms $S_{ab}(\tau)$ are replaced by their sample counterparts:
\begin{equation}
    \hat{S}_{ab}(\tau)=\frac{1}{N}\sum_{k=1}^{N-\tau} x_{a}(k)x_{b}(k+\tau)
\end{equation}
\section{Computer Experiments}\label{sec:computer-experiments}
\paragraph*{Illustration on copula.} Our goal is to generate colored multivariate non-Gaussian time-series, whose marginals are Gaussian to make the problem more difficult. With this goal, we chose to use Archimedean copula for their ease to generate in dimension $p>2$.
\begin{definition}\label{def:d-archimedean}
A $p$-dimensional copula $\mathcal{C}_\rho$ is called Archimedean if it allows the
representation:
\begin{equation}
    \mathcal{C}_{\rho}(u) = \psi(\psi^{-1}(u_1) + \dots+ \psi^{-1}(u_p)), u \in [0,1]^{p}
\end{equation}
\end{definition}
for some Archimedean generator $\psi$ and its inverse $\psi^{-1}$:
$(0,1] \to [0,\infty$).

\begin{figure}[htb]
\begin{minipage}[b]{0.48\linewidth}
  \centering
  \centerline{\includegraphics[width=\textwidth]{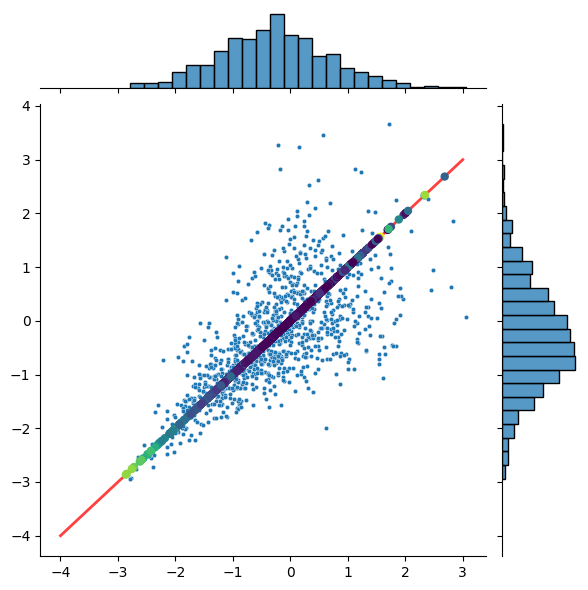}}
  %\vspace{1.5cm}
  \centerline{(b) Clayton $\varphi = \frac{\pi}{4}$}\medskip
\end{minipage}
\hfill
\begin{minipage}[b]{.48\linewidth}
 \centering
  \centerline{\includegraphics[width=\textwidth]{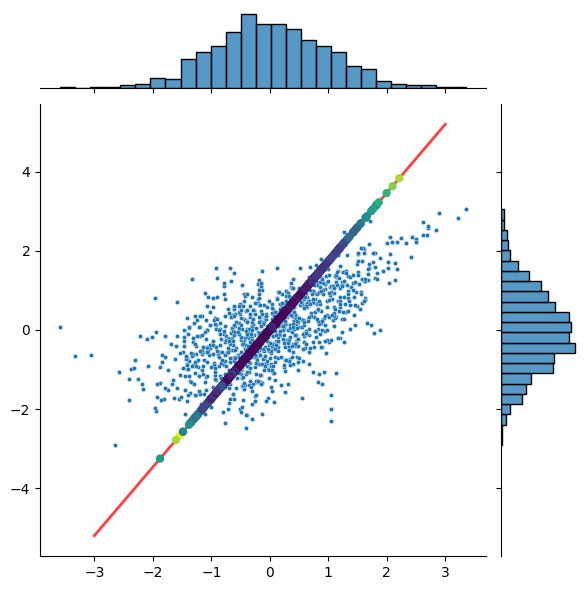}}
  \centerline{(a) Gumbel $\varphi = \frac{\pi}{6}$}\medskip
\end{minipage}

\caption{ Two examples of projecting \textbf{bivariate} realizations (in blue) onto the direction in red defined by the angle $\varphi$. Realizations are generated using Clayton copula (on the left), and Gumbel copula (on the right). The distribution of canonical marginals are both Gaussian as illustrated by histograms but the bivariate distribution is clearly not Gaussian.}
\label{fig:1D_project}
\begin{minipage}[b]{0.48\linewidth}
  \centering
  \centerline{\includegraphics[width=1.3\textwidth]{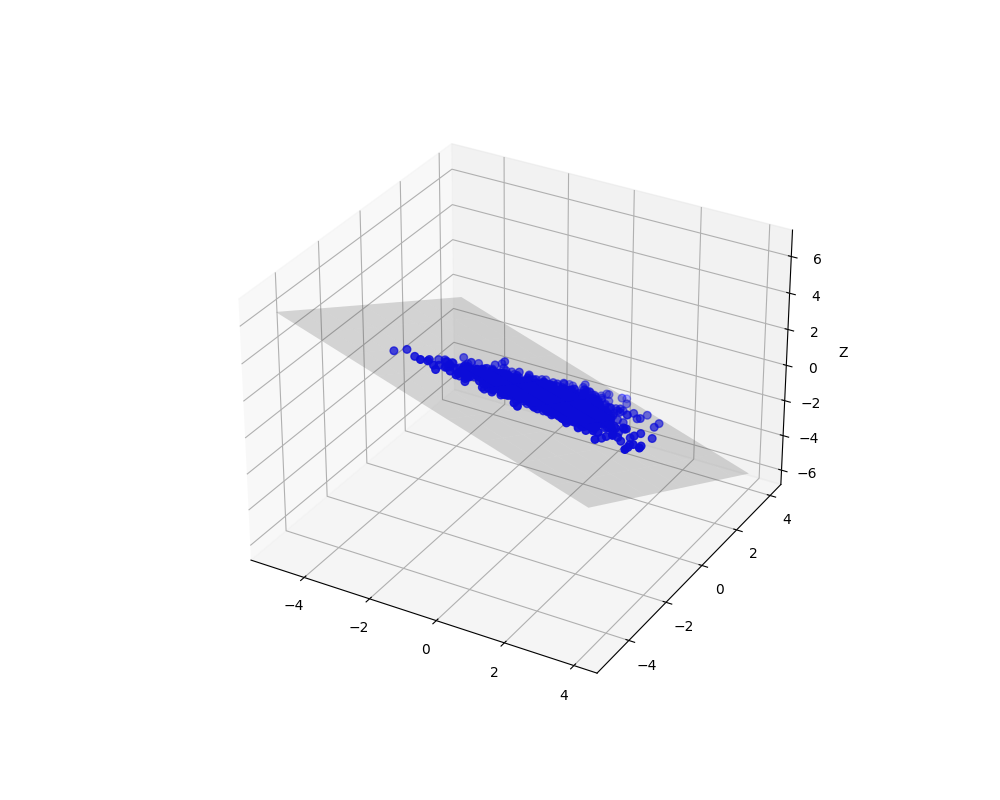}}
  \centerline{(c) Clayton  }\medskip
\end{minipage}
\hfill
\begin{minipage}[b]{0.48\linewidth}
  \centering
  \centerline{\includegraphics[width=1.3\textwidth]{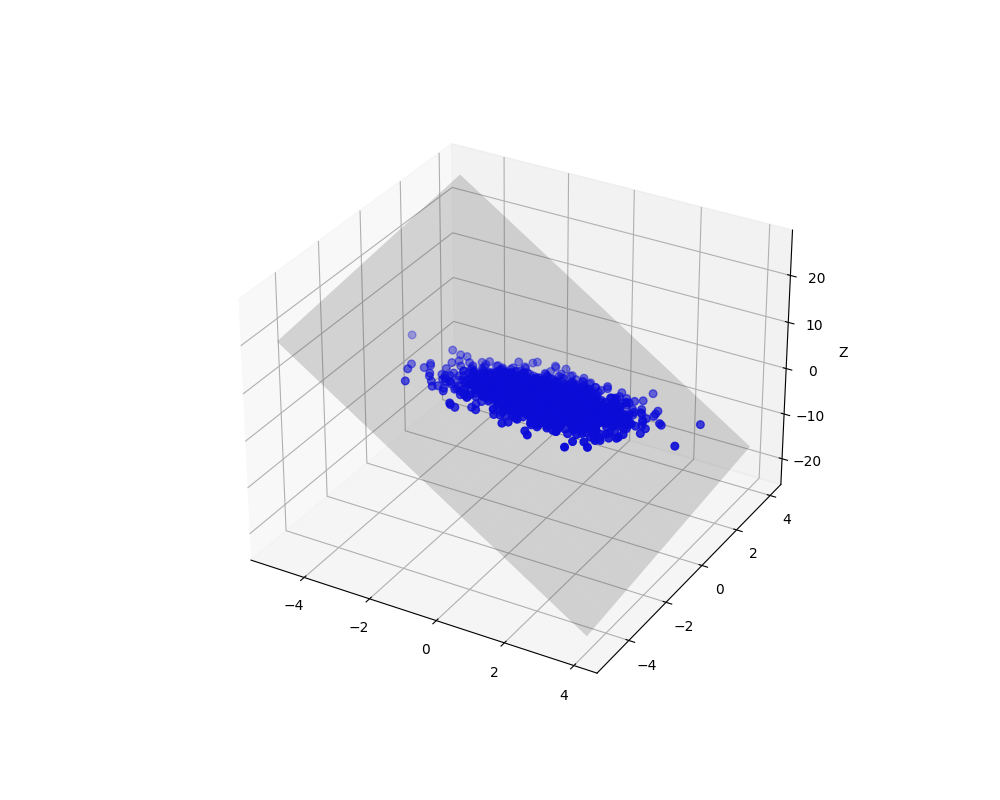}}
  \centerline{(d) Gumbel }\medskip
\end{minipage}
\caption{Two examples of projecting \textbf{trivariate} realizations onto the plane (in light grey). Realizations are generated using Clayton copula (on the left), and Gumbel copula (on the right). The distribution of the two-dimensional projection is clearly not Gaussian.}
\label{fig:2D_project}
\end{figure}

The parameter $\rho$ of the copula is related to Kendall rank correlation coefficient. Thus, it controls the spatial dependence between variables. In order to introduce time dependency between samples, an AR filter is applied on each marginal before constructing copula (this preserves normality). This leads to the following algorithm:

\paragraph*{Sampling an Archimedean copula.}
\begin{enumerate}
    \item Sample i.i.d $\eta_{i} \sim \mathcal{N}(0,1)$, $i \in \{1,\hdots,p\}$
    \item Correlate $\eta_{i}$'s using a first order auto-regressive filter:
    \begin{eqnarray}
    y_{i}(n)&=&0.8 y_{i}(n-1)+\eta_{i}(n) \nonumber
    \end{eqnarray}
    Note that the first $n_{\text{drop}}=1000$ samples are dropped to alleviate start-up effects ($y_{i}(0)=\eta_{i}(0)$).
    \item Transform $u_i= \Phi(y_i)$ for $i \in \{1,\hdots,p\}$, where $\Phi$ denotes the cdf of the Gaussian distribution. Note that $u_i$'s are uniform on $[0,1]$.
    \item Sample $V \sim \mathcal{LS}^{-1}(\psi)$ where $\mathcal{LS}^{-1}$ denotes the inverse Laplace-Stieltjes transform of $\psi$
    \item Return ($u'_{1}, u'_{2},\dots,u'_{p}$), where $u'_{i}=\psi(-\log(u_i)/V)$
    \item Transform $u^{\prime}_i$  to obtain Gaussian standard marginals as the following: \begin{equation}
        x_{i}(n) = \Phi(u^{\prime}_i(n))
    \end{equation}
\end{enumerate}
The above algorithm is a slight modification to the one  due  to   Mashall, Olkin (1988)\cite{Hofe08:csda}. In the remainder  of this paper, we precisely use  Gumbel ($\rho=5$) and  Clayton ($\rho=2$) copula.
% Below is an example of how to insert images. Delete the ``\vspace'' line,
% uncomment the preceding line ``\centerline...'' and replace ``imageX.ps''
% with a suitable PostScript file name.
% -------------------------------------------------------------------------
\paragraph*{Low-dimensional projection.}\noindent We study the performance of the proposed test statistic on a low-dimensional (either $1$ or $2$) projection of the initial $p$-variate data. For a given copula $\mathcal{C}_\rho$, we carried out the following simulations:
\begin{itemize}
    \item Given one set of bivariate observations ($x_{1}(n),x_{2}(n)$)
    of total length $N=1000$, they are projected $M=5000$  times onto the arbitrary vector $\vect{u}$ with coordinates ($\sin(\varphi),\cos(\varphi)$). $\varphi$ is sampled from a uniform distribution on $[0,\pi]$ denoted $\mathcal{U}(0,\pi)$. \textbf{Fig.\ref{fig:1D_project}} shows an illustrative example with two copulas.
    \item Given one set of trivariate observations of total length $N=1000$, the points are projected arbitrarily $M=5000$ times onto the plane defined by two angles $\theta \sim \mathcal{U}(-\frac{\pi}{2}, \frac{\pi}{2})$ (the angle between the $z$ axis and the new plane) and $\varphi \sim \mathcal{U}(0,\pi)$ (measured between the $x$ axis and the vector $\vect{u}$ inside the plane). \textbf{Fig. \ref{fig:2D_project}} gives two illustrative examples of this procedure.
\end{itemize}
For each projection, we measure the $p$-values returned by the test statistic. Each $p$-value is compared with the level of the test $\alpha$; if inferior to it the test rejects the null hypothesis of normality.
The empirical rejection
rates, summarized in the following tables are computed as $\frac{\textnormal{Number of Rejections}}{M}$.

\section{Main results}\label{sec:mainresults}
We report the empirical rejection rate of each test statistic mentioned in \textbf{Tables} \textbf{\ref{tab:scalar-timedep-project}, \ref{tab:scalar-timeIndep-project}, \ref{tab:two-arb-projections}} and \textbf{\ref{tab:bivariate-timedep-project}} and comment on the results for each scenario mentioned in the top left cell of each table.
The results are reported for two significance levels $\alpha = 5\%$ and $\alpha=10\%$.
\pagebreak

\paragraph*{Scalar projection} 
\begin{itemize}
    \item $\hat{B}_{1}$ and $\hat{B}_{1,i.i.d}$ perform very poorly when used on arbitrary one-dimensional projections of the Gumbel copula. The test power does not surpass $25\%$. 
    \item For the Clayton copula, whose tails are asymmetric, the test has a better power than the Gumbel copula. Although this observation is less demonstrative, we keep those results to further compare them with the bivariate test statistic.
    \item Since we only use first-order auto-regressive filters, there is no substantial difference in the performance of $\hat{B}_{1}$ compared to $\hat{B}_{1,i.i.d}$; this comparison is not of interest to us, because the bias induced by using tests assuming independence on colored processes has already been observed and studied in the literature \cite{Gass75:biom} and \cite{Moor82:as}. \\
    However, it is interesting to compare \textbf{Tables} \textbf{\ref{tab:scalar-timedep-project}} and \textbf{\ref{tab:scalar-timeIndep-project}}; we see that the overall performance of the test statistics tends to decrease when marginals are \textbf{time-correlated}.
    \item \textbf{Table \ref{tab:two-arb-projections}} shows performances obtained with $\hat{B}_2$ when applied to colored processes. Contrary to $\hat{B}_1$, performances do not decrease with time-correlation. 
    Furthermore, the power of the 2-D test based on $\hat{B}_2$ is not affected by a rotation in the plane (implemented by two scalar projections onto two orthogonal axes). This is illustrated by  \textbf{Table \ref{tab:two-arb-projections}}, which reports the results averaged over 5000 random rotations. 
\end{itemize}
\paragraph*{Bivariate projection}
\begin{itemize}
\item One would expect the same problem of misdetections to occur when projecting trivariate observations sampled from Gumbel copula. Yet, in \textbf{Table \ref{tab:bivariate-timedep-project}} we show that the joint normality, even on a low representation of the data, is able to detect the non-Gaussianity of the process.
\end{itemize}
%\begin{figure}[htb]
%\begin{minipage}[b]{0.48\linewidth}
%  \centering
%  \centerline{\includegraphics[width=\textwidth]{figures/Histogram_arbitrary_projection_Gumbel5_notid_5000iter.png}}
%  \centerline{(a) Gumbel copula $\theta=5$ }\medskip
%\end{minipage}
%\hfill
%\begin{minipage}[b]{0.48\linewidth}
%  \centering
%  \centerline{\includegraphics[width=\textwidth]{figures/Histogram_arbitrary_projection_NOTID_CLAYTON2_2d_to_3d_5000iter.png}}
%  \centerline{(b) Clayton copula $\theta=2$}\medskip
%\end{minipage}
%\caption{Histogram of the $p$-values of $5000$ realizations of the test applied on projections with arbitrary angles $\rho \sim \mathcal{U}(0,\pi)$}
%\end{figure}
\begin{table}[htb]
\centering
\begin{tabular}{|l|llll|}
\hline
\multirow{2}{*}{\begin{tabular}[c]{@{}l@{}}Time \\ correlated\end{tabular}} & \multicolumn{2}{l|}{$\hat{B}_{1}$}                                     & \multicolumn{2}{l|}{$\hat{B}_{1,i.i.d}$}          \\ \cline{2-5} 
                                                                                     & \multicolumn{1}{l|}{$\alpha=5\%$} & \multicolumn{1}{l|}{$\alpha=10\%$} & \multicolumn{1}{l|}{$\alpha=5\%$} & $\alpha=10\%$ \\ \hline
Gumbel                                                                    & 0.1250                            & 0.1328                             & 0.1242                            & 0.1316        \\ \hline
Clayton                                                               & 0.661  & 0.72
&            0.652       & 0.713\\ \hline
\end{tabular}
\caption{Empirical rejection rates of the test applied to a one-dimensional projection of a bivariate process with time-correlated Gaussian marginals.}
\label{tab:scalar-timedep-project}
\end{table}
\begin{table}[htb]
\centering
\begin{tabular}{|l|llll|}
\hline
\multirow{2}{*}{\begin{tabular}[c]{@{}l@{}}Time\\independent\end{tabular}} & \multicolumn{2}{l|}{$\hat{B}_{1}$}                                     & \multicolumn{2}{l|}{$\hat{B}_{1,i.i.d}$}          \\ \cline{2-5} 
                                                                                        & \multicolumn{1}{l|}{$\alpha=5\%$} & \multicolumn{1}{l|}{$\alpha=10\%$} & \multicolumn{1}{l|}{$\alpha=5\%$} & $\alpha=10\%$ \\ \hline
Gumbel                                                                     & 0.2134                            & 0.2510                             & 0.2082                            & 0.2406        \\ \hline
Clayton                                                                       & 0.717                        & 0.76                             & 0.701                           & 0.752     \\ \hline
\end{tabular}
\caption{Empirical rejection rates of the test applied to a one-dimensional projection of a bivariate process with independent (in time) standard normal marginals.}
\label{tab:scalar-timeIndep-project}
\end{table}
\begin{table}[!htb]
\centering
\begin{tabular}{|l|ll|}
\hline
\multirow{2}{*}{\begin{tabular}[c]{@{}l@{}} Arbitrary 2-D\\ projections  \end{tabular}} & \multicolumn{2}{l|}{$\hat{B}_{2}$}                \\ \cline{2-3} 
                                                                                     & \multicolumn{1}{l|}{$\alpha=5\%$} & $\alpha=10\%$ \\ \hline
Gumbel                                                               & 0.9516                            & 0.9574        \\ \hline
Clayton                                                                   & 0.9701                            & 0.9882        \\ \hline
\end{tabular}
\caption{Empirical rejection rates of the test applied jointly to arbitrary 2-D projections.}
\label{tab:two-arb-projections}
\end{table}
\begin{table}[!htb]
\centering
\begin{tabular}{|l|ll|}
\hline
\multirow{2}{*}{\begin{tabular}[c]{@{}l@{}}Time-correlated\\ marginals\end{tabular}} & \multicolumn{2}{l|}{$\hat{B}_{2}$}                \\ \cline{2-3} 
                                                                                     & \multicolumn{1}{l|}{$\alpha=5\%$} & $\alpha=10\%$ \\ \hline
Gumbel                                                                    & 0.9492                            & 0.9556        \\ \hline
Clayton                                                                 & 0.8540                           & 0.87         \\ \hline
\end{tabular}
\caption{Empirical rejection rates of the test applied to the two-dimensional projection of a trivariate process with time-correlated marginals.}
\label{tab:bivariate-timedep-project}
\end{table}
%\vfill
%\pagebreak


\vfill\pagebreak
\section{Concluding remarks}
This study demonstrates, on one hand, that testing the joint normality of a two-dimensional projection yields a noticeable increase in the power of the test to detect departure from joint normality, even in the most pathological scenario of the Gumbel copula. On the other hand, when data are additionally time-correlated, the overall power of scalar tests tends to decrease. By assuming both spatial and temporal dependence, our bivariate test stands out from other existing multivariate tests that assume independence.

Future studies will be carried out to validate the performance of this statistic on real higher dimensional data.

\bibliographystyle{IEEEbib}
\bibliography{refs}
\end{document}